\documentclass[aps,prl,twocolumn,superscriptaddress]{revtex4-2}

\usepackage{amsmath}
\usepackage{amsfonts}
\usepackage{amssymb}
\usepackage{dcolumn}
\usepackage[english]{babel}
\usepackage{epsfig}
\usepackage{graphics}
\usepackage{xcolor}

\begin{document}

\title{Observation of Phase Space Dynamics of Inverted Harmonic Oscillator}

\author{Georgi Gary Rozenman}
\affiliation{Raymond and Beverly Sackler School of Physics $\&$ Astronomy, Faculty of Exact 
Sciences, Tel Aviv University, Tel Aviv 69978, Israel}
\affiliation{School of Electrical and Computer Engineering, Iby and Aladar Fleischman Faculty of Engineering, 
Tel Aviv University, Tel Aviv 69978, Israel}
\affiliation{Department of Mathematics, Massachusetts Institute of Technology, Cambridge, Massachusetts, USA}

\author{Maxim A. Efremov}
\affiliation{German Aerospace Center (DLR), Institute of Quantum Technologies, 89081 Ulm, Germany}

\author{Wolfgang P. Schleich}
\affiliation{Institut f\"ur Quantenphysik and Center for
Integrated Quantum Science and Technology ($\it IQST$), Universit\"at Ulm, 89081 Ulm, Germany}
\affiliation{Hagler Institute for Advanced Study at Texas A$\&$M University, 
Texas A$\&$M AgriLife Research, Institute for Quantum Science and Engineering (IQSE), and Department of Physics and Astronomy, 
Texas A$\&$M University, College Station, TX 77843-4242, USA}

\author{Lev Shemer}
\affiliation{School of Mechanical Engineering, Faculty of Engineering, Tel Aviv University, Tel Aviv 69978, Israel}

\author{Ady Arie}
\affiliation{School of Electrical and Computer Engineering, Iby and Aladar Fleischman Faculty of Engineering, 
Tel Aviv University, Tel Aviv 69978, Israel}

\date{\today}

\begin{abstract}
{We have experimentally realized a parabolic potential barrier for surface gravity water waves. The analogy of the resulting wave equation to the Schr\"odinger equation of the inverted harmonic oscillator (IHO) enables us to study the propagation of quantum mechanical wave packets with different average energy in this iconic scattering model. We have observed a clear boundary in the phase-space dynamics --- the separatrix --- separating wave packets with energy below the maximum of the IHO potential from those above it. In the former case the wave is blocked, whereas in the latter it is transmitted. We have also measured the variation in momentum in this process.}
\end{abstract}


\maketitle

\noindent{\it Introduction.--}{The harmonic oscillator is arguably {\it the} fundamental problem of quantum mechanics and at the very heart of many phenomena in atomic, molecular, nuclear, and particle physics \cite{feynman1965feynman}. Its lesser-known counterpart, the {\it inverted} harmonic oscillator (IHO), exhibits a parabolic {\it barrier} rather than a {\it binding} potential and is therefore central to quantum mechanical scattering \cite{barton1986quantum}. In the present Letter, we realize an IHO in the Schr\"odinger-like wave equation for surface gravity water waves, and 
observe the associated phase-space dynamics \cite{schleich2011quantum}.

The IHO displays many characteristic features that make it an interesting model system in numerous fields ranging from cell-biology via sociophysics \cite{weidlich2012concepts} to economics \cite{mantegna1999introduction} and finances. In quantum mechanics it can be considered a paradigmatic illustration of phenomena taking place in a quasi-infinite space. Three examples may suffice to illustrate this point.  

The effect of quantum mechanical tunneling through a barrier \cite{feynman1965feynman} finds numerous applications in modern technology, ranging from the transistor via the Josephson effect to coupled waveguide resonators. The parabolic barrier provided by the IHO is the most elementary example \cite{Kemble1935} and provides deeper insight \cite{balazs1990wigner,heim2013tunneling} into the non-classical phenomenon of tunneling.

In the field of quantum optics \cite{schleich2011quantum} the IHO also plays a prominent role. Energy eigenstates in the classically forbidden regime provide an elementary explanation for the gain in a laser \cite{glauber2007}. Moreover, they are crucial for the understanding of the phenomenon of difference-frequency resonance \cite{vazquezChampneys2010}, the quantum amplification by superradiant emission of radiation \cite{svidzinsky2013}, and the measurement of trajectories without uncertainties \cite{moller2017}.

Moreover, an IHO and a black hole share many common features \cite{subramanyan2021physics} when viewed through the lens of phase space rather than space-time.
Indeed, already the classical IHO displays a separatrix in phase space which translates in quantum mechanics into a logarithmic phase singularity   \cite{bib:Ullinger2022} which is a key feature of the Hawking radiation \cite{hawking1974black} emitted by a black hole. A recent experiment \cite{rozenman2024} using surface gravity water waves studied the phase-space event horizon of an IHO. Related analogue experiments have measured stimulated Hawking emission \cite{Weinfurtner2011} and the scattering of co-current surface waves on an analogue black hole \cite{Euve2020}.



Motivated by its rich dynamics \cite{SchmiedmayerIHO2026}, we implement the IHO using classical surface-gravity water waves \cite{Mei, rozenman2019quantum}, which obey a wave equation analogous to the Schr\"odinger equation \cite{Fu2015a}. In our system, the parabolic barrier is realized by a homogeneous, time-dependent water current. This temporal, rather than spatial realization follows from the interchange of space and time in the effective Schr\"odinger-like equation \cite{rozenman2023observation}.

We focus on the evolution of Gaussian wave packets in this classical arrangement to gain insight into the corresponding quantum problem. In particular, we map out the phase-space dynamics and demonstrate the stopping power of the IHO.

}

\begin{figure}
  \includegraphics[width=0.80\columnwidth]{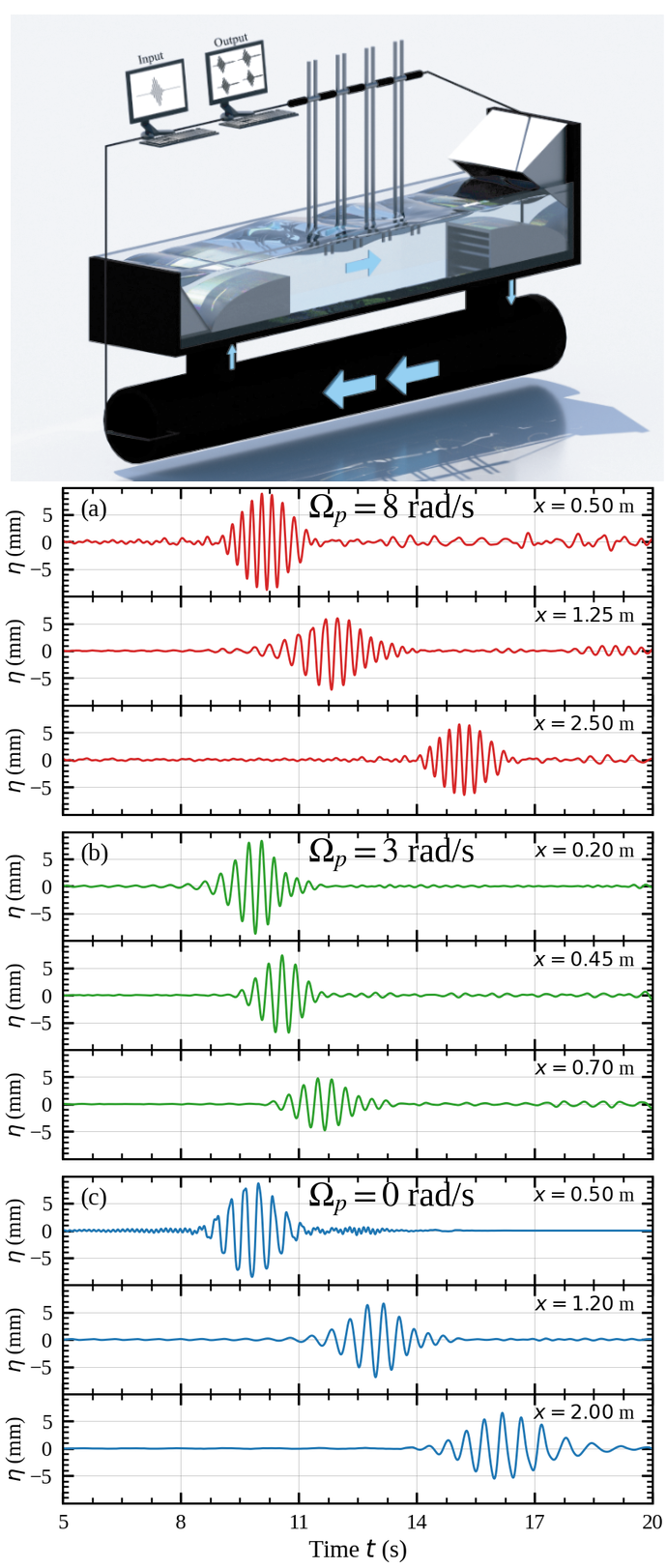}\\
  \caption{\label{fig:scheme1}
  Scattering of Gaussian wave packets in an inverted harmonic oscillator (IHO): the experimental setup and the measured surface-elevation data. Surface gravity water waves in a tank obey a wave equation analogous to the quantum-mechanical Schr\"odinger equation, but with time and space interchanged. The parabolic barrier is realized by a water current (blue arrows) with a time dependence of an inverted parabola. Wave packets are launched at the right by a computer-controlled wave maker and recorded by four gauges along the propagation direction; the two screens display a typical input packet (top) and the measured, time-dependent surface elevation $\eta=\eta(t)$ (bottom). We consider (a)~transmission above the barrier (red, $E_0>0$), (b)~motion along the separatrix (green, $E_0=0$), and (c)~reflection below the barrier (blue, $E_0<0$), which we analyze in Figs.~\ref{fig:phasespace} and \ref{fig:trajectories}. For each case, the three panels show the surface elevation recorded by the wave gauges at three stages of the experiment --- at the beginning, in the middle (at the maximum of the IHO potential), and at the end.}
\end{figure}
{

\noindent{\it A quadratic barrier for surface gravity water waves.--}
In many aspects, the time evolution of a wave function in quantum mechanics is analogous to that of paraxial optical beams \cite{rodrigues2022bright}, 
surface gravity water wave pulses \cite{rozenman2019amplitude,weisman2021diffractive,rozenman2022periodic,rozenman2023observation,Chabchoub2013} and 
underwater acoustic beams \cite{bar2015observation}.
For surface gravity water waves with low steepness, propagating in a homogeneous flow with a quadratic time-dependence, the wave equation
\begin{equation}
 \label{Schroedinger-equation}
 -i\frac{\partial A}{\partial\xi}=-\frac{\partial^2 A}{\partial\tau^2}- \Omega^2 \tau^{2} A
\end{equation}
for the normalized complex-valued amplitude envelope $A\equiv A(\tau, \xi)$ in the {\it co-moving frame} has a form \cite{Mei} similar to the one-dimensional time-dependent Schr\"odinger equation of a particle in "an inverted harmonic potential" $-\Omega^2 \tau^{2}$ with a constant "frequency" $\Omega$.

However, the roles of time and space are interchanged. Indeed, the scaled
dimensionless variables $\xi$ and $\tau$ are related to the propagation
coordinate $x$ and the time $t$ by
$\xi \equiv \varepsilon^{2} k_{0} x$ and
$\tau \equiv \varepsilon \omega_{0} \left( x / c_{g_{0}} - t \right)$.
The carrier wave number $k_{0}$ and angular frequency $\omega_{0}$ define
the group velocity $c_{g_{0}} = \omega_{0} / (2k_{0})$ and obey the
deep-water dispersion relation $\omega_{0}^{2} = g k_{0}$, which is
characteristic of freely propagating surface-gravity waves in the absence
of an external current, i.e.\ for a vanishing flow velocity $U(t)=0$.
Here, $g$ denotes the gravitational acceleration.
The wave steepness parameter $\varepsilon \equiv k_{0} a_{0}$ is assumed
to be sufficiently small to ensure that
Eq.~(\ref{Schroedinger-equation}) remains free of nonlinear terms
\cite{Mei, Rozenman2020}. Here $a_0$ is the maximum wave amplitude.

For $U(t)\neq 0$, i.e. in the presence of a spatially homogeneous but
time-dependent current, the carrier frequency becomes time dependent and
follows the modified deep-water dispersion relation \cite{baos2020}
$\omega(t) = U(t) k_{0} + \sqrt{g k_{0}}$.
Here, $U(t)$ denotes the velocity of the external
homogeneous current and is defined as
$U(t) = U_{0} + c\,(t - t_{p})^{2}$, where $t_{p} = 4.0\,\mathrm{s}$ is the
time at which the current reaches its maximum, $U_{0} = 0.1354\,\mathrm{m/s}$, and $c = -0.00841\,\mathrm{m/s^{3}}$.
In our experiments, we have used the carrier wave frequency $\omega_{0}=17\,\mathrm{rad/s}$ and the wave vector $k_{0}=29.5\,\mathrm{1/m}$ satisfies the deep-water condition $k_{0}h>\pi$, with $h$ the water depth. 
The amplitude is $a_{0}=10^{-3}\,\mathrm{m}$ and the small steepness, $\varepsilon=0.0295$, guarantees the validity of the linear Schr\"odinger equation, Eq.~(\ref{Schroedinger-equation}).


The effective potential $-\Omega^2 \tau^{2}$ in Eq. (\ref{Schroedinger-equation}) is determined by
the derivative $\left(\partial \Phi/\partial\tau\right)|_{Z=0}$ of the external dimensionless velocity potential 
$\Phi\equiv \phi/(\omega_0a_0^2)$ at the surface given by the dimensionless vertical coordinate $Z=0$ 
with $Z\equiv \varepsilon k_0 z$ \cite{Mei}. The current velocity follows from the velocity potential $\phi$ by $U=\partial\phi/\partial x$, as for an irrotational flow~\cite{LandauLifshitz1987}. Hence, we can create the potential $\Omega^{2} \tau^{2}\equiv 4\varepsilon\left(\partial \Phi/\partial\tau\right)|_{Z=0}$ by 
an externally operating water pump with the appropriate time-dependent current. We emphasize that the inverted-parabolic form $-\Omega^2 \tau^{2}$ represents the imposed barrier only in the vicinity of its maximum: since the externally driven current cannot grow without bound, the effective potential is necessarily truncated and brought back to zero far from the barrier top, and our measurements are confined to the region around the maximum where the parabolic approximation holds.

\begin{figure}[h]
  \includegraphics[width=0.87\columnwidth]{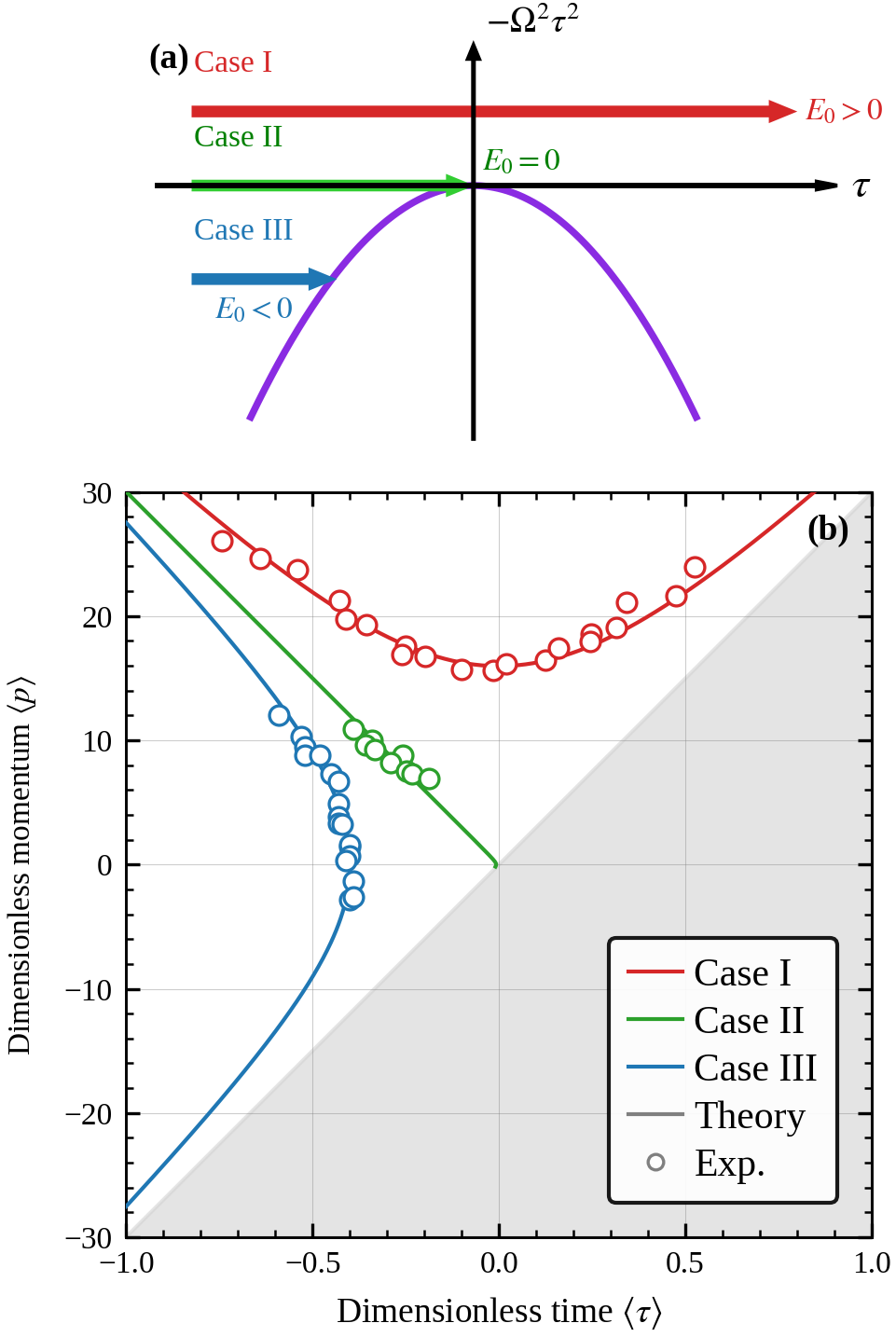}\\
  \caption{\label{fig:phasespace}
  Experimental observation of the phase-space trajectories of an IHO using Gaussian surface-gravity water wave packets.
(a) The three scattering situations correspond to different energies $E_0$,
leading to transmission (red curves), stopping (green curves), or
reflection (blue curves) of the wave packet.
(b) Phase space of the IHO represented by the
three trajectories associated with the energies shown in (a).
Open circles represent experimental values obtained by combining the
measurements shown in Figs.~\ref{fig:trajectories}(a) and
\ref{fig:trajectories}(b); the solid curves follow from
Eqs.~(\ref{coordinate-mean-value-Gaussian}) and
(\ref{momentum-mean-value-Gaussian}) upon elimination of $\xi$.
The shaded area indicates the domain of phase space not accessible to
wave packets approaching from the left, and the diagonal boundary
defines the separatrix. Here, in Figs.~\ref{fig:scheme1} and \ref{fig:trajectories}, we have chosen the values $\Omega_p = 8$, $3$, and $0$~rad/s for Cases~I, II, and III (red, green, and blue), respectively.
}
\end{figure}


\textit{Results. -}
To observe the phase space dynamics of Gaussian wave packets in the IHO, we have conducted a series of experiments with surface gravity water waves moving in the so-prepared time-dependent water flow using the experimental facility shown in Fig. \ref{fig:scheme1}. The water waves are generated by a computer-controlled wave maker at one end of the tank which is $5\,{\rm m}$ long, $0.4\,{\rm m}$ wide, and $0.2\,{\rm m}$ deep, and are measured by capacitance-type wave gauges in the regions of interest. A wave-energy-absorbing beach is placed at the other end of the water tank. To eliminate the effect of reflections from the beach, precise measurements of the water surface elevation are carried out by wave gauges at distances not exceeding $4.0\,{\rm m}$ from the wave maker, i.e. 1.0 meter away from the absorbing beach.

\begin{figure}[h]
  \includegraphics[width=1.0\columnwidth]{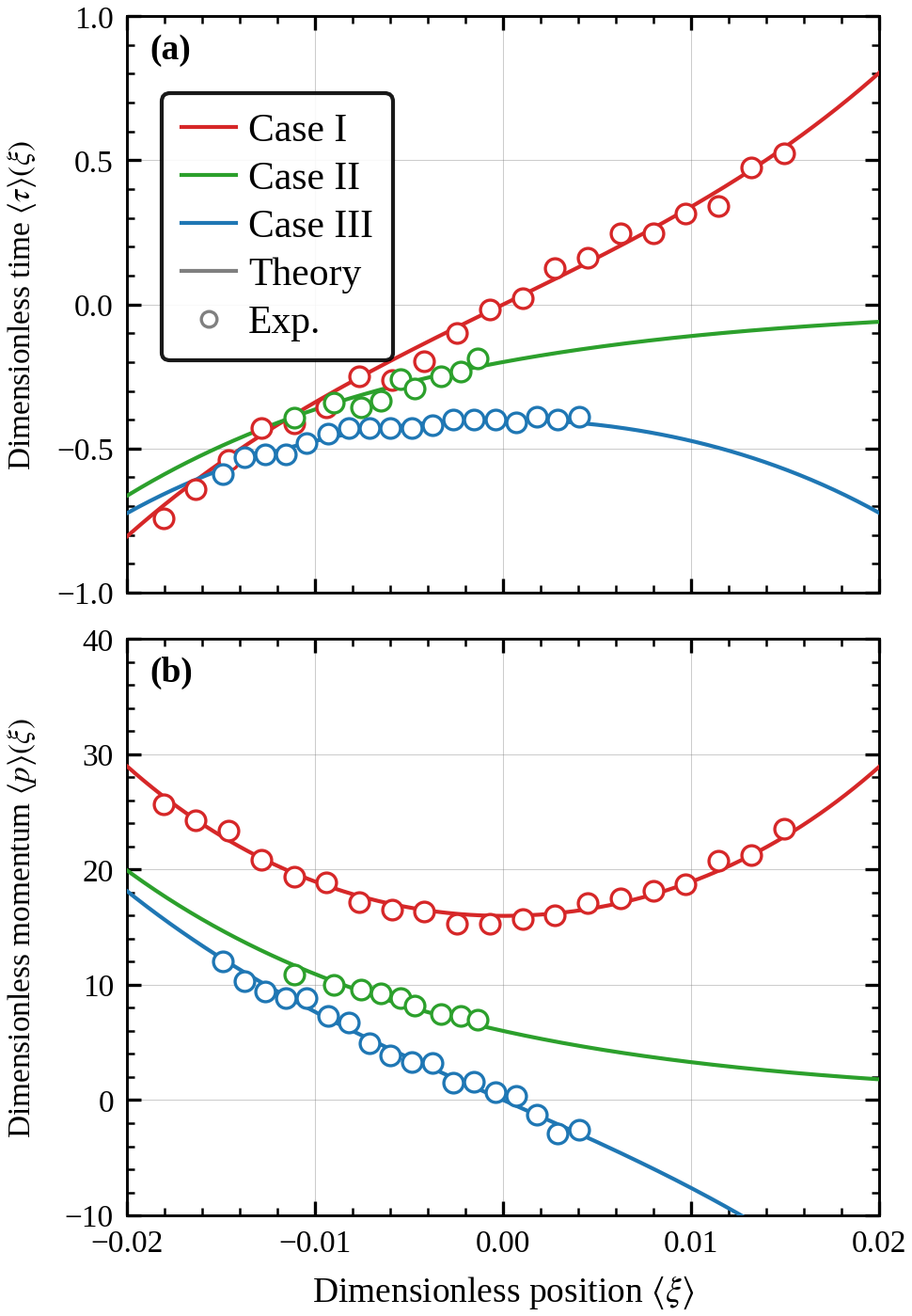}\\
  \caption{\label{fig:trajectories}
  Comparison between the experimentally observed (open circles) and theoretically predicted (solid lines) trajectories of analog time coordinate $\langle \tau  \rangle (\xi) $ (a), and analog momentum $\langle p \rangle (\xi)$ (b), for the three values of $\Omega_{p}$ given in Fig.~\ref{fig:scheme1}.}
\end{figure}

The complex amplitude envelope $A\equiv\left|A\right| \exp\left(i\varphi_{A}\right)$ determines the variation in time and space of the surface elevation 
\begin{equation}
 \label{eta}
  \eta(t,x)\equiv {\rm Re}\left[a_0 A(t,x)e^{i(k_0 x-\omega_0 t)}\right]
\end{equation}
including the carrier wave.

The Gaussian envelope of the temporal surface elevation generated by the wave maker at $x=0$ corresponds to the initial condition
\begin{equation}
A^{(G)}(\tau,\xi=0)=\exp\left[-\left(\frac{\tau-\tau_{0}}{\Delta\tau}\right)^2 - i p_{0} \tau \right],
\label{A-initial-Gaussian}
\end{equation}
where $\tau_0=-\varepsilon\omega_0 t_0$, $\Delta\tau=\varepsilon\omega_0\Delta t$, and $p_0=\Omega_p/(\varepsilon\omega_0)$. Here $\Omega_p$ and $p_0$ are the effective momentum and its dimensionless counterpart; Eqs.~(\ref{eta}) and (\ref{A-initial-Gaussian}) show that they correspond to a shift of the carrier frequency. 

Under the evolution governed by Eq.~(\ref{Schroedinger-equation}), the packet remains Gaussian, and its center follows the trajectory
\begin{equation}
\langle\tau\rangle(\xi)=\tau_0\cosh(2\Omega\xi)+\frac{p_0}{\Omega}\sinh(2\Omega\xi).
\label{coordinate-mean-value-Gaussian}
\end{equation}

Depending on the classical energy $E_0=p_0^2-\Omega^2\tau_0^2$, which coincides with the expectation value $\langle H\rangle$ of the energy, this trajectory represents a transmission above the barrier, $E_0>0$, a motion along the separatrix at zero energy, $E_0=0$, or reflection below the barrier, $E_0<0$, as displayed in Fig. 2(a). The corresponding measured surface-elevation profiles for these three cases are shown in Figs.~\ref{fig:scheme1}(a)--(c).

To observe these dynamics experimentally, we measure $\eta=\eta(t,x)$ at 20 spatial positions. Next, we reconstruct the complex amplitude $A(\tau,\xi)$ from Eq.~\eqref{eta}, with the help of the Hilbert transform, as detailed in the appendix of Ref.~\cite{rozenman2024}. From $A$ we then determine the expectation value
\begin{equation}
\langle \tau \rangle(\xi)=N^2\int_{-\infty}^{\infty} d\tau \, A^{*}(\tau,\xi)\,\tau\,A(\tau,\xi),
\label{position-mean-value_integral}
\end{equation}
defining the trajectory of the analog time coordinate, where
\[
N^2=\left(\int_{-\infty}^{\infty} d\tau\, |A(\tau,\xi)|^2\right)^{-1}.
\]

Figure~3(a) presents the measured trajectories $\langle\tau\rangle(\xi)$ for the three values of $\Omega_p$ of Fig.~\ref{fig:scheme1}, together with fits, given by the corresponding solid lines, Eq.~(\ref{coordinate-mean-value-Gaussian}). From these fits we obtain the values $\Omega=29.80\pm0.65$, $29.72\pm0.77$, and $30.87\pm0.69$, respectively, for the frequency of the IHO, resulting in the mean value $\bar{\Omega}=30.14\pm0.40$.

We next measure the corresponding analog momentum, defined as
\begin{equation}
\langle p\rangle (\xi)=N^2\int_{-\infty}^{\infty} d\tau \, A^{*}(\tau,\xi)\,i\frac{\partial}{\partial\tau}\,A(\tau,\xi),
\label{momentum-mean-value-def}
\end{equation}
and derive in the End Matter the relation
\begin{equation}
\langle p\rangle(\xi)=p_0\cosh(2\Omega\xi)+\Omega\tau_0\sinh(2\Omega\xi).
\label{momentum-mean-value-Gaussian}
\end{equation}

\begin{figure}
  \includegraphics[width=1.0\columnwidth]{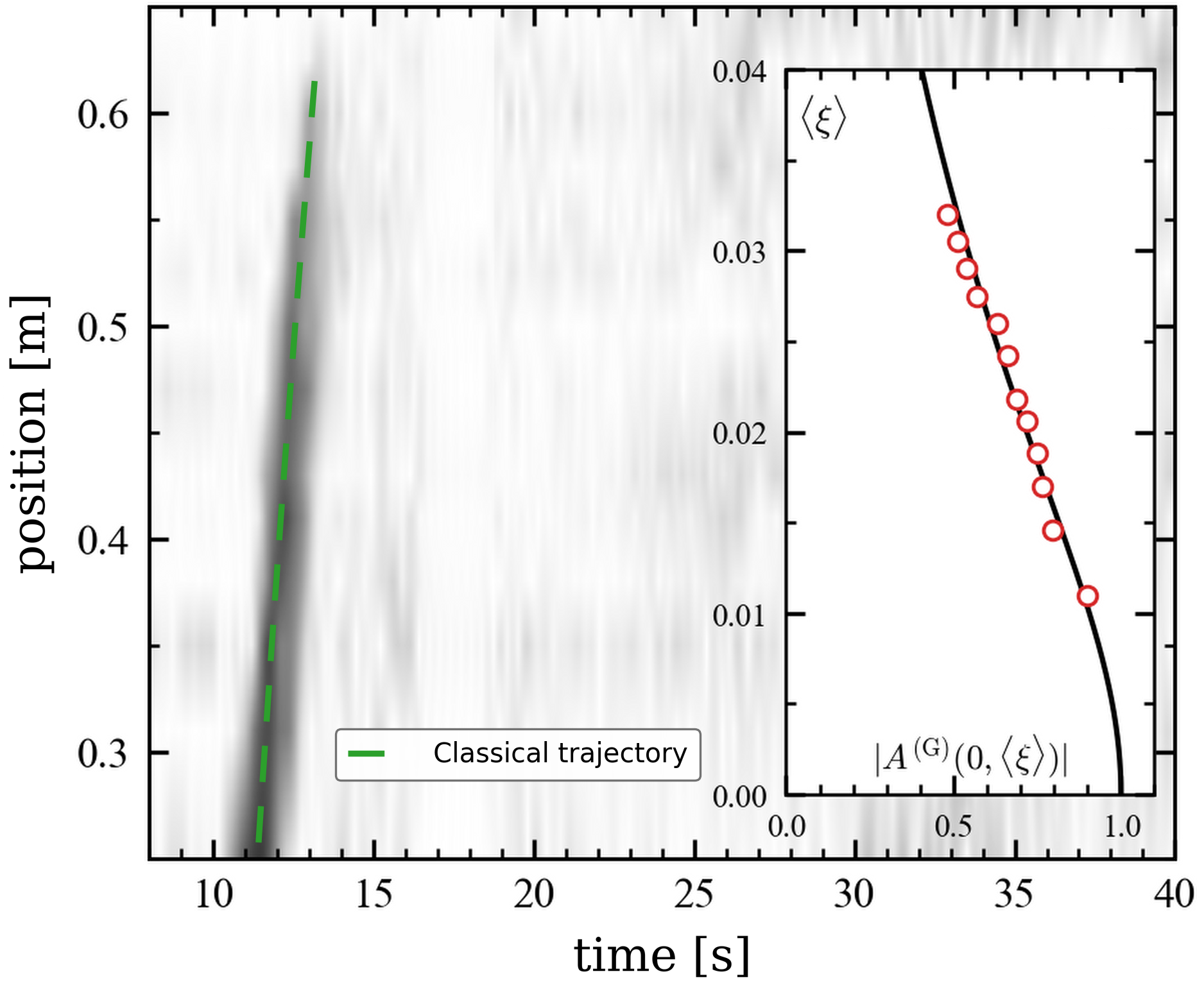}\\
  \caption{\label{fig:black}
  Observation of the propagation of a Gaussian surface gravity water wave packet with an average energy equal to the top of the barrier, represented by a density plot of the absolute value $|A^{(G)}(x,t)|$ in the laboratory frame. Dark and light shades indicate large and small values of $|A^{(G)}(x,t)|$. The wave packet follows the classical trajectory given by Eq.~(\ref{coordinate-mean-value-Gaussian}) transformed into the laboratory frame and indicated by the green dashed line. As the wave packet propagates, it spreads and its amplitude at the top of the barrier decays, as shown in the inset: open red circles are the experimental values and the black solid line is the analytical prediction obtained by evaluating the Gaussian envelope of Eq.~(\ref{amplitude-Gaussian-endmatter}) at $\tau=0$, i.e.
$|A^{(G)}(0,\xi)|=[\Delta\tau/\Delta\tau(\xi)]^{1/2}\exp\{-[\langle\tau\rangle(\xi)/\Delta\tau(\xi)]^{2}\}$,
with $\langle\tau\rangle(\xi)$ and $\Delta\tau(\xi)$ given by Eqs.~(\ref{coordinate-mean-value-Gaussian}) and (\ref{width-endmatter}). The wave packet disappears into the noise before reaching the location of the potential maximum, i.e.\ the top of the barrier at analog time $\tau=0$, which on the separatrix is approached only asymptotically (see inset).
}
\end{figure}

In Fig. 3(b) we present the measured momentum analog $\langle p \rangle (\xi)$ for the same three different values of $\Omega_p$, as in Fig. 3(a), by open circles. Next, we fit these data by the corresponding theoretical prediction, Eq. (\ref{momentum-mean-value-Gaussian}), and again extract the effective IHO frequency $\Omega$. We find the values $\Omega=30.91 \pm 0.24$ (red), $\Omega=28.93 \pm 0.54$ (green) and $\Omega=30.24 \pm 0.71$ (blue), resulting in a mean value $\bar{\Omega}=30.55 \pm 0.21$, which is in good agreement with the result obtained from the trajectories of the analog time coordinate $\langle \tau \rangle (\xi)$. 

Finally, we reconstruct classical {\it phase-space} trajectories by combining the experimentally obtained trajectories $\langle\tau\rangle=\langle\tau\rangle(\xi)$ and $\langle p \rangle=\langle p \rangle(\xi)$ presented in Fig. \ref{fig:trajectories} and eliminating $\xi$. In Fig. \ref{fig:phasespace}(b), we show the so-obtained phase-space trajectories by open circles, and compare them to the ones resulting from eliminating $\xi$ in Eqs. (\ref{coordinate-mean-value-Gaussian}) and (\ref{momentum-mean-value-Gaussian}) and depicted by solid lines.

A particularly interesting case emerges when the average energy of the wave packet is identical to the top of the barrier, that is $E_0=0$, as shown by the green arrow in Fig.~\ref{fig:phasespace}(a). In this case, the wave packet propagates along the separatrix in phase space indicated in Fig.~\ref{fig:phasespace}(b) in green.

In Fig. \ref{fig:black}  we show the resulting dynamics in the laboratory frame and note that, due to the spreading of the wave packet, its amplitude is below the noise level of the wave gauges when it reaches the location of the maximum of the IHO.
Moreover, a white corridor seems to also suggest a reduction of the noise in the neighborhood of the top of the barrier. A plausible explanation is that the rapid, current-induced stretching of the surface field near the barrier top dilutes both the wave packet and the ambient background fluctuations, so that the locally recorded noise is reduced in this region, consistent with the separatrix dynamics at zero average energy.

In our experiments we have not observed reflections of portions of the wave packet when scattering from the potential. In the IHO no quantum reflections occur for positive energy values \cite{Friedrich2002, Bestle1995}. Other reflections always occur when parts of the Wigner function of the wave packet cover domains above and below the separatrix; this phenomenon takes place when the characteristic length of the potential is small compared to the size of the wave packet \cite{barton1986quantum}. However, our experiments have not been in this regime.


\noindent{\it Conclusion.--}
We have experimentally realized an IHO for surface gravity water waves by means of a homogeneous, quadratically time-dependent current, and have used this platform to map out its phase-space dynamics. Three concrete results follow from our measurements: (i) For Gaussian wave packets with positive, vanishing, or negative classical energy $E_0$, we have reconstructed the trajectories of the analog time coordinate $\langle\tau\rangle(\xi)$ and analog momentum $\langle p\rangle(\xi)$, and demonstrated the three characteristic regimes of the IHO --- transmission, motion along the separatrix, and reflection --- in quantitative agreement with the analytical hyperbolic predictions, Eqs.~(\ref{coordinate-mean-value-Gaussian}) and (\ref{momentum-mean-value-Gaussian}). (ii) The two independent extractions of the value of the IHO frequency from the coordinate and momentum trajectories agree within their statistical uncertainties, confirming the consistency of the analogy. (iii) For the marginal case $E_0=0$, we have observed the predicted amplitude decay of the wave packet at the top of the barrier, in quantitative agreement with the Gaussian envelope $|A^{(G)}(0,\xi)|$ derived in the End Matter.

\noindent{\it Outlook.--}
The IHO is a paradigmatic model in several branches of physics. It contains an elementary parabolic barrier of quantum scattering and tunneling \cite{Kemble1935,barton1986quantum,balazs1990wigner,heim2013tunneling}, it is at the very heart of the phase-space description of black-hole horizons and Hawking radiation due to a logarithmic singularity at the separatrix \cite{subramanyan2021physics,bib:Ullinger2022,hawking1974black,rozenman2024}, and it appears in fields well outside of physics \cite{weidlich2012concepts,mantegna1999introduction}. Our experimental realization of an IHO paves the way for numerous new developments. Three examples may suffice to illustrate possible research directions: (i) Probe the quantum-reflection regime, where the characteristic length of the barrier becomes comparable to the wave-packet width, and partial reflection above the barrier appears \cite{barton1986quantum,Friedrich2002,Bestle1995}. (ii) Measure directly the logarithmic phase singularity at the separatrix, which sets the analog Hawking spectrum \cite{bib:Ullinger2022,rozenman2024}, and (iii) implement Kostin-type time-dependent control \cite{kostin1972schrodinger,losert2023kostin} to bring wave packets to rest. We emphasize that we can extend many of these ideas to optical, acoustic, and matter-wave systems \cite{rodrigues2022bright,bar2015observation}, opening a cross-platform avenue for studying parabolic-barrier physics with direct relevance to quantum scattering, gravitational analogs, and non-equilibrium dynamics in classical and quantum systems.


\qquad

We thank Matthias Zimmermann, Alona Maslennikov and Anatoliy Khait for fruitful discussions, help, and support, and Tamir Ilan for technical support and advice. This work is supported by DIP, the German-Israeli Project Cooperation (AR 924/1-1, DU 1086/2-1) supported by the DFG, the Israel Science Foundation (Grant No. 969/22, No. 508/19), and the Israel Ministry of Science, Technology and Space (Grant No. 3-12473). W.P.S. is grateful to Texas A$\&$M University for a Faculty Fellowship at the Hagler Institute for Advanced Study at the Texas A$\&$M University as well as to the Texas A$\&$M AgriLife Research. The research of the IQST is financially supported by the Ministry of Science, Research and Arts Baden-W\"urttemberg.


\section*{End Matter}
\subsection*{Expectation value of analog momentum}

In this section we summarize the derivation of the expectation value $\langle\hat{p}\rangle$ of the analog momentum operator
\begin{equation}
\hat{p}\equiv i\frac{\partial}{\partial\tau}.
\end{equation}  
Starting from the Gaussian wave packet, Eq.~(\ref{A-initial-Gaussian}), we write the complex amplitude envelope in the form
\begin{equation}
A^{(G)}(\tau,\xi)=|A^{(G)}(\tau,\xi)|e^{i\varphi_A^{(G)}(\tau,\xi)},
\end{equation}
and find the expression
\begin{equation}
\begin{split}
\langle \hat{p}\rangle =\;&
N^2\int_{-\infty}^{\infty} d\tau \,
A^{*}(\tau,\xi)\hat{p}A(\tau,\xi) \\
=\;&
-N^2\int_{-\infty}^{\infty} d\tau \,
|A(\tau,\xi)|^2
\left(\frac{\partial \varphi_A^{(G)}}{\partial \tau}\right).
\end{split}
\label{momentum-mean-value-endmatter}
\end{equation}

The phase $\varphi_A^{(G)}$ of the Gaussian envelope $A^{(G)}$ consists of the sum \cite{rozenman2019amplitude}
\begin{equation}
\varphi_A^{(G)}=\varphi_G+\varphi_{cm}+\varphi_K,
\end{equation}
of the Gouy
\begin{equation}
\varphi_G\equiv
\frac{1}{2}
\arctan\!\left[
\frac{2}{\Omega\Delta\tau^2}\tanh(2\Omega\xi)
\right],
\label{gouy-phase-endmatter}
\end{equation}
the center-of-mass
\begin{equation}
\varphi_{cm}\equiv
\frac{\Omega\Delta\tau^2}{2\tanh(2\Omega\xi)}
\left[
\frac{\tau-\langle\tau\rangle(\xi)}{\Delta\tau(\xi)}
\right]^2,
\label{cm-phase-endmatter}
\end{equation}
and the propagation
\begin{equation}
\varphi_K\equiv
-\frac{\Omega}{2}
\frac{(\tau^2+\tau_0^2)\cosh(2\Omega\xi)-2\tau\tau_0}
{\sinh(2\Omega\xi)}
\label{phase-Gaussian-endmatter}
\end{equation}
phases.

With the help of the explicit expression \cite{rozenman2019amplitude}
\begin{equation}
|A^{(G)}(\tau,\xi)|
=
\left[\frac{\Delta\tau}{\Delta\tau(\xi)}\right]^{1/2}
\exp\!\left\{
-\left[
\frac{\tau-\langle\tau\rangle(\xi)}{\Delta\tau(\xi)}
\right]^2
\right\},
\label{amplitude-Gaussian-endmatter}
\end{equation}
for $A^{(G)}$, with the propagated width
\begin{equation}
\Delta\tau(\xi)
=
\Delta\tau\,\sqrt{\cosh^{2}(2\Omega\xi)
+\left[\frac{2}{\Omega\Delta\tau^{2}}\right]^{2}\!\sinh^{2}(2\Omega\xi)}\,,
\label{width-endmatter}
\end{equation}
and the average trajectory
\begin{equation}
\langle\tau\rangle(\xi)
=
\tau_0\cosh(2\Omega\xi)
+
\frac{p_0}{\Omega}\sinh(2\Omega\xi),
\end{equation}
we note that only the propagator phase $\varphi_K$ contributes to the integral in Eq.~(\ref{momentum-mean-value-endmatter}).

Indeed, according to Eq.~(\ref{gouy-phase-endmatter}), the Gouy phase $\varphi_G$ is independent of $\tau$, so that its derivative with respect to $\tau$ vanishes. Moreover, the center-of-mass phase $\varphi_{cm}$, Eq.~(\ref{cm-phase-endmatter}), is symmetric in the difference $\tau-\langle\tau\rangle$, so that its derivative with respect to $\tau$ is antisymmetric; since $|A^{(G)}|^2$ is symmetric in $\tau-\langle\tau\rangle$, the corresponding integral over $\tau$ vanishes as well.

When we substitute the propagation phase $\varphi_K$ defined by Eq.~(\ref{phase-Gaussian-endmatter}), together with the expression, Eq.~(\ref{amplitude-Gaussian-endmatter}), for $|A^{(G)}|^2$, into Eq.~(\ref{momentum-mean-value-endmatter}), perform the differentiation with respect to $\tau$ and carry out the integration over $\tau$, we arrive at the formula
\begin{equation}
\langle p\rangle(\xi)
=
p_0\cosh(2\Omega\xi)
+
\Omega\tau_0\sinh(2\Omega\xi),
\label{momentum-mean-value-Gaussian-endmatter}
\end{equation}
which is the expression used in the main text to fit the experimental data and reconstruct the phase-space trajectories.

\newpage

\end{document}